\documentclass[prb,aps,twocolumn,showpacs,preprintnumbers,superscriptaddress]{revtex4-1}
\usepackage{latexsym,epsfig,amssymb,amsfonts,amsmath,graphicx,color,bm,times,hyperref,amstext,epstopdf,braket,ulem} 

\usepackage{soul}
\usepackage{color}

\begin{document}

\title{Anderson localization effects near the Mott metal-insulator transition}

\author{Helena Bragan\c{c}a}

\author{M. C. O. Aguiar}
\affiliation{Departamento de F\'isica, Universidade Federal de Minas Gerais, 
Belo Horizonte, MG, Brazil}

\author{J. Vu\v{c}i\v{c}evi\'c}

\author{D. Tanaskovi\'c}
\affiliation{Scientific Computing Laboratory, Institute of Physics Belgrade,
University of Belgrade, Pregrevica 118, 11080 Belgrade, Serbia}

\author{V. Dobrosavljevi\'c}
\affiliation{Department of Physics and National High Magnetic Field Laboratory,
Florida State University, Tallahassee, Florida 32306}

\date{\today}

\begin{abstract}

The interplay between Mott and Anderson routes to localization in disordered interacting systems gives rise to
different transitions and transport regimes. Here, we investigate the phase diagram at finite temperatures using
dynamical mean field theory combined with typical medium theory, which is an effective theory of the
Mott-Anderson metal-insulator transition. We mainly focus on the properties of the coexistence region associated with
the Mott phase transition. For weak disorder, the coexistence region is found to be similar as in the clean case. 
However, as we increase disorder Anderson localization effects are responsible for shrinking the coexistence region and 
at sufficiently strong disorder (approximately equal to twice the bare bandwidth) it drastically narrows, the critical temperature $T_c$ abruptly goes to zero, and we observe a phase transition in the absence of a coexistence of the metallic and insulating phases. In this regime, the effects of interaction and disorder are found to be of comparable importance for charge localization.

\end{abstract}







\maketitle

\section{Introduction}

Mott mechanism of localization\cite{Mott} is an emergent phenomenon in which a large local 
Coulomb repulsion suppresses double occupation, which prevents charge transport in a half-filled system.
Strongly correlated electron materials, such as transition metal oxides 
\cite{ImadaRMP,Exp1,Exp2,Exp3} and some organic salts,\cite{KagawaNature2005,MerinoPRL2008,Exp4,Exp5,Sasaki2012} 
exhibit a Mott metal-insulator transition due to the effectively strong Coulomb repulsion that 
exists between electrons occupying a narrow valence band. Below the critical temperature $T_c$, 
this transition is of first-order and one observes a region where metal and insulator 
coexist.\cite{Exp2,Exp3,KagawaNature2005,Exp4}

The presence of disorder also leads to localization of electron wave functions, a phenomenon 
known as Anderson localization.\cite{Anderson,LeeRMP1985} In this case the energetic mismatch between 
neighboring sites prevents charge transport in the lattice. These two mechanisms of localization 
- Mott and Anderson -  combine in nontrivial ways, sometimes reducing, sometimes enhancing 
each other's effects.
Recently, the interplay between interaction and disorder has received much attention, mainly 
through three different perspectives. First, due to the investigation of the many-body 
localization,\cite{ManyBody} a novel paradigm arose for understanding localization in 
disordered and interacting quantum systems at non-zero temperature. Second, very 
recently, models of disordered and interacting systems have been simulated with cold atoms 
in optical lattices.\cite{OPL1,OPL2} 
Finally, the disorder and the effective interaction strength can be systematically tuned
by doping,\cite{Exp1, Exp3, Exp5, Exp6, Exp7} or even X-ray irradiation.\cite{Sasaki2012,Analytis}

Over the last few decades considerable progress in the description of strongly correlated 
materials and Mott metal-insulator transition (MIT) has been achieved through 
dynamical mean field theory (DMFT).\cite{GeorgesRMP1996} In this method, a lattice model of interacting electrons is mapped 
to the Anderson impurity model with a conduction bath which needs to be calculated self-consistently.
To describe disorder, the simplest treatment is within the coherent potential approximation (CPA).\cite{Economou2005}
The CPA can be easily combined with the DMFT,\cite{JanisPRB1992,Ulmke1995,LaadPRB2001,ByczukPRL2003, ByczukPRB2004,Carol1,Radonjic2010,Poteryaev2015} 
by considering an ensemble of impurities surrounded by an average bath, which is the same for each electron.
This approach thus does not describe the spatial fluctuations associated with the Anderson localization. Near the Anderson transition the distribution of the local density of states (DOS) changes from Gaussian to log-normal,\cite{SchubertPRB2010,VladNew} implying that its arithmetic average value does not provide a proper description of the system. The typical medium theory (TMT)\cite{TMT} provides a simple method which is able to effectively describe the Anderson localization. The central quantity in TMT is the typical density of states, defined as the geometric average of the local DOS,
\cite{Janssen} which plays the role of the order parameter for the Anderson localization. The TMT method was carefully tested for the noninteracting system,\cite{TMT,EkumaJPCM2014,EkumaPRB2014} and it was successfully applied to the interacting case within the TMT-DMFT approach,\cite{Vollhardt} elucidating the full nonmagnetic phase 
diagram for the disordered half-filled Hubbard model and  the precise nature of the Mott-Anderson critical point.\cite{Carol2}
The TMT-DMFT approach also allows for a spin-dependence analysis of the DOS, which enables one to include effects of long-range magnetic order in disordered and interacting systems.\cite{Mag}

In this paper, we perform the first TMT-DMFT calculation at finite temperatures.
We explore the entire nonmagnetic phase diagram with a particular focus on the effects of disorder
on the Mott metal-insulator coexistence region. We carefully compare the TMT-DMFT and CPA-DMFT results
with the goal of precisely determining the Anderson localization effects, described only within the former method.
We find that the TMT-DMFT coexistence region is at comparatively lower values
of the interaction $U$, while the critical temperature $T_c$ is higher than in CPA-DMFT.
The width of the coexistence region, however, quickly decreases
with disorder. At disorder strength $W \sim 2B$, where $B$ is
the bandwidth in the clean noninteracting system, TMT-DMFT predicts $T_c$ to abruptly go to zero, as opposed to the CPA-DMFT
solution where the coexistence region asymptotically shrinks to a single point as disorder is increased to infinity.
In the regime $W \gtrsim 2B$ the MIT takes place at $U \approx W$, which makes Anderson and Mott mechanisms to become equally important for the properties of the system.

The paper is organized as follows. 
In section~\ref{II} we briefly present the TMT-DMFT method for the solution of the disordered Hubbard model, 
and the $(U,W)$ phase diagram is shown is section~\ref{III}. Sections \ref{IV} and \ref{V} show details of the
metal-insulator transition in the presence of weak, moderate, and strong disorder. Section \ref{VI} contains conclusions.

\section{TMT-DMFT method}\label{II}

We consider the Hubbard model with random site energies, given by the Hamiltonian
$$
  H= -t \sum_{\langle ij \rangle \sigma} (c_{i \sigma}^{\dagger}c_{j \sigma} + \mathrm{H.c.}) + U \sum_i n_{i \uparrow} n_{i \downarrow} + \sum_{i \sigma} (\varepsilon_i - \mu) n_{i \sigma},  
$$
where $c^{\dagger}_{i \sigma}$ ($c_{i \sigma}$) creates (destroys) an electron with spin 
$\sigma$ on site $i$, $n_{i \sigma}= c^{\dagger}_{i \sigma} c_{i \sigma}$, 
$t$ is the hopping amplitude for nearest neighbor sites, $U$ is the on-site repulsion
and $\varepsilon_i$ is the random on-site energy, which follows a uniform distribution 
$P(\varepsilon)$ of width $W$, centered in $\varepsilon_i =0$. We study the half-filled particle-hole symmetric lattice by setting the chemical potential $\mu$ equal to $U/2$.
In general, transition metal oxides and organic salts described by the Hubbard model can exhibit both antiferromagnetic and paramagnetic Mott insulating phases. 
In this work, we focus on the paramagnetic solution which is present even at zero temperature in frustrated lattices.

Within TMT-DMFT, the lattice model describing a disordered correlated 
system is mapped onto an ensemble of single-impurity problems, corresponding 
to sites with different values of the on-site energy, each being embedded 
in a typical effective medium that needs to be calculated self-consistently. 
The TMT-DMFT self-consistent procedure can be summarized as 
follows:\cite{TMT,Carol2} 
By considering an initial guess for the (typical) bath $\Delta(\omega)$ 
surrounding the impurities, we solve an ensemble of impurity problems, which
give us local Green's functions
$G(\omega,\varepsilon_i)$ from which local spectra
$\rho(\omega, \varepsilon_i)=- \frac{1}{\pi} \mbox{Im}G(\omega, \varepsilon_i)$
are obtained. The typical DOS is then calculated by the geometric average of the local spectra, 
$$
\rho_{typ}(\omega)=exp\left[\int d \varepsilon P(\varepsilon) \ln \rho(\omega, \varepsilon)\right],
$$
and the typical Green's function is obtained through the Hilbert transform,
$G_{typ}(\omega)= \int_{-\infty}^{\infty} d \omega \frac{\rho_{typ}(\omega')}{\omega- \omega'}$.
For lattices with semicircular DOS, $\rho_0(\omega)=\frac{4}{\pi B}\sqrt{1-\left(\frac{2 \omega}{B}\right)^2}$,
in the clean non-interacting limit (Bethe lattice with infinite 
coordination number), the  self-consistent loop is closed by calculating a new bath according to 
$\Delta(\omega)=t^2G_{typ}(\omega)$. 
To solve the single-impurity problems, in this work
we use the iterative perturbation theory (IPT) on the real axis.\cite{Kajueter,Potthoff} In this case we do not need analytic continuation. This is an important advantage of this method since the TMT self-consistency relation is based on the local DOS.

\section{Phase diagram}\label{III}

\begin{figure}
\includegraphics[scale=0.35]{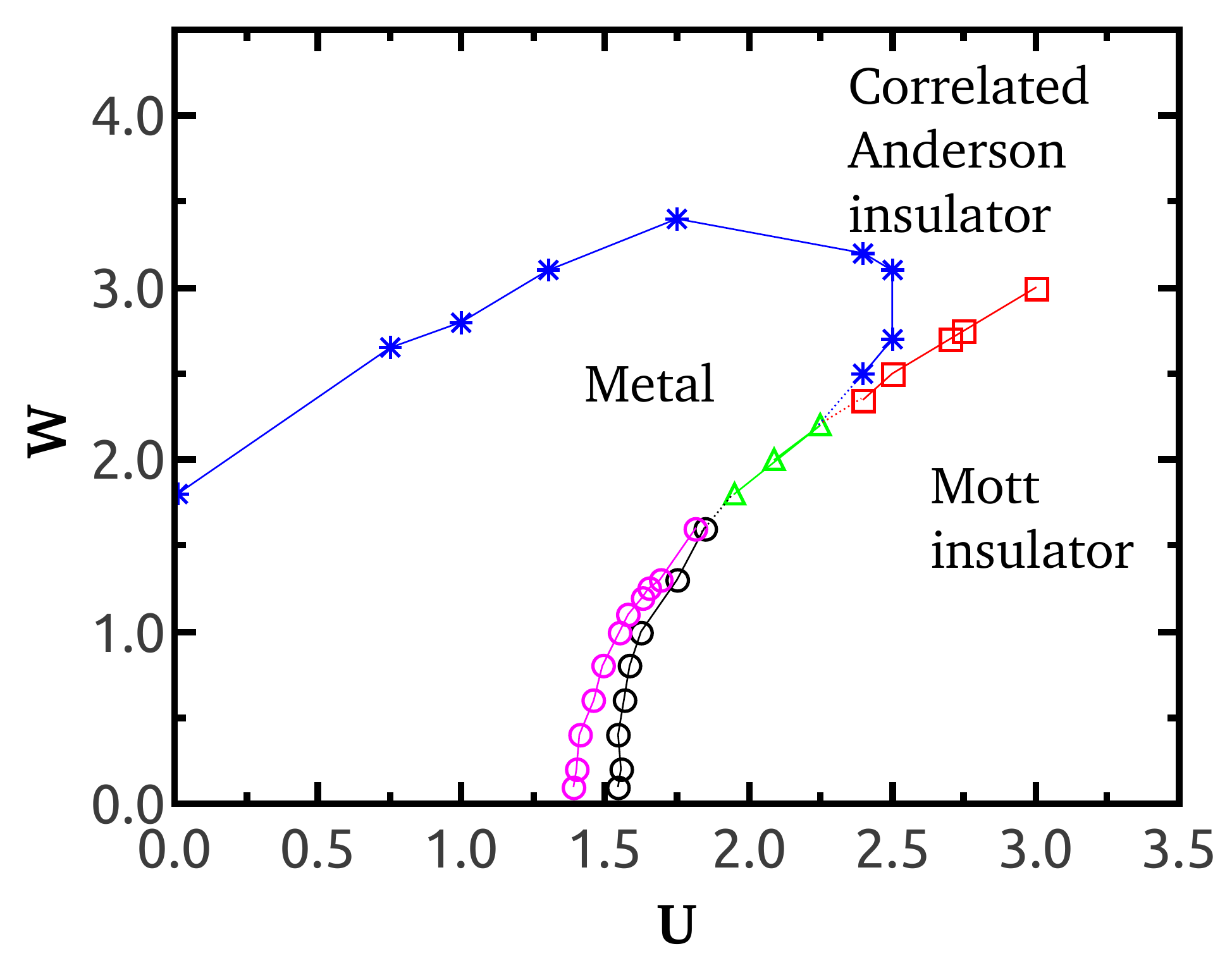}
\caption{(Color online) $(U , W)$ phase diagram obtained within TMT-DMFT for the 
disordered Hubbard model at $T=0.008$. The description of the different symbols/colors 
used is given in the text.} 
\label{fig1}
\end{figure}

Fig.~\ref{fig1} presents the TMT-DMFT phase diagram of 
the disordered Hubbard model obtained at a small 
temperature, $T=0.008$. Here and throughout the paper we define the non-interacting
bandwidth $B=4t$ as the unit of energy. In the phase diagram, the black and pink circles 
correspond to the metallic and the insulating spinodal lines of the first-order Mott phase transition; 
these two lines delimit the metal-insulator coexistence region. The green triangles 
indicate a transition between a metal and a Mott insulator in the absence of 
a well defined coexistence region (see Section \ref{V} for details), while the blue 
stars indicate 
a transition between a metal and a correlated Anderson insulator. Finally, the red squares 
correspond to a crossover between the two insulators, which takes place at $W \approx U$. 

\begin{figure}
\includegraphics[width=2.8 in]{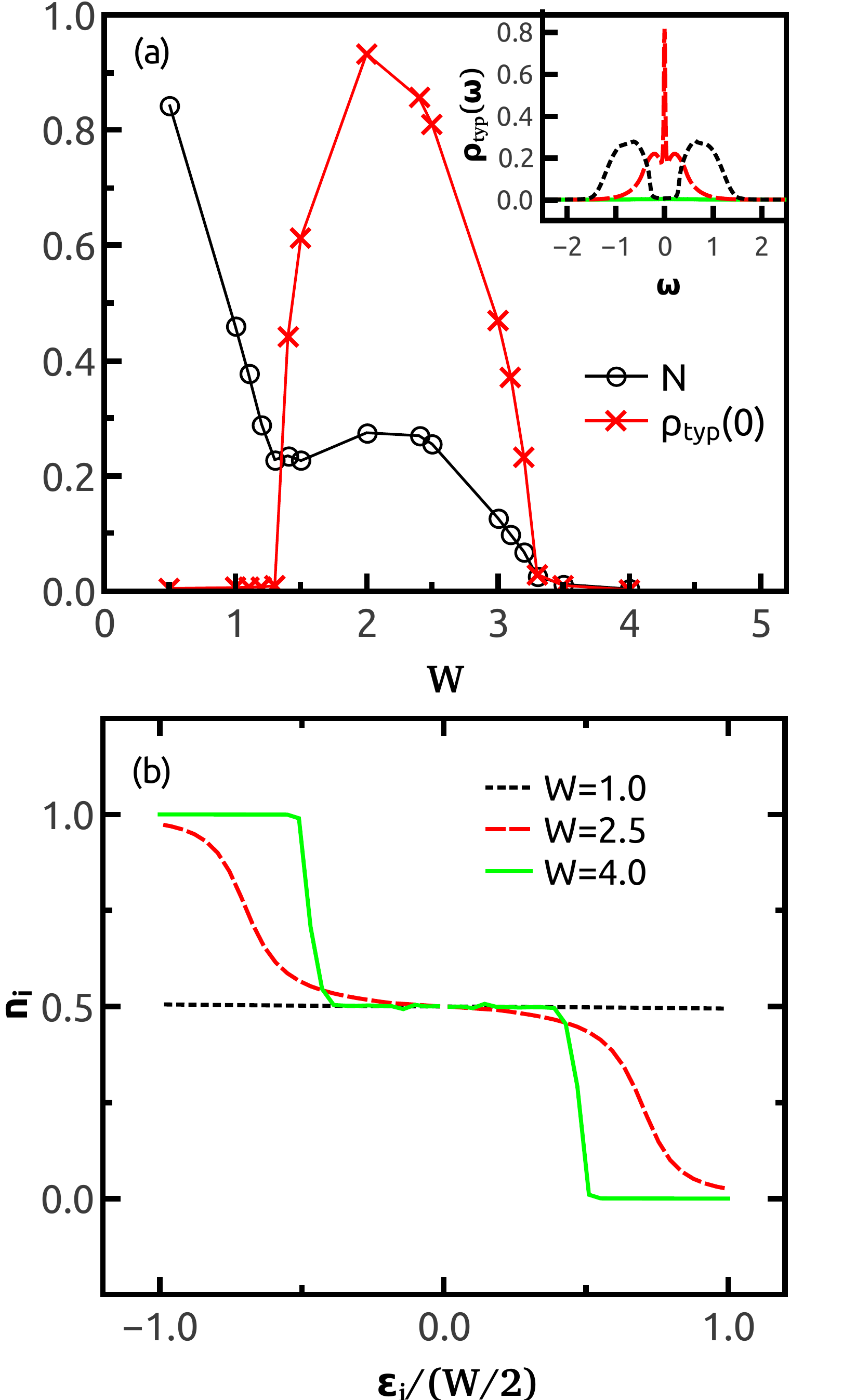}
\caption{(Color online) TMT-DMFT results for $U=1.75$ and $T=0.008$. According to the 
phase diagram of Fig.~\ref{fig1}, as disorder $W$ increases,
the system goes from the Mott insulating phase to the metallic phase and finally to the 
Anderson insulating phase. These transitions are identified (and the phase diagram is 
built) by looking at the behavior of the quantities shown in the two panels of the present 
figure: (a) frequency integrated typical DOS, $N$, and 
typical DOS at the Fermi level, $\rho_{typ}(0)$, as a function of $W$, and (b) site occupation 
per spin as a function of the site energy, normalized by the disorder distribution width, $W$.
The inset shows an example of the typical DOS in the metallic phase (red long-dashed line), 
as well as in the Mott (black dashed line) and the Anderson (green solid line) insulating
phases.} 
\label{fig2} 
\end{figure}

To differentiate the phases and build the phase diagram, we have analyzed the behavior of
the typical DOS at the Fermi level ($\rho_{typ}(0)$), the frequency integrated typical 
DOS ($N$) and the site occupation as a function of the on-site energy. As an example, these 
quantities are presented in Fig.~\ref{fig2} for the particular case of $U=1.75$ and $T=0.008$. 
For this set of parameters, as disorder $W$ increases, the system goes 
from the Mott insulator to the Anderson insulator, crossing an intermediate metallic phase
(see Ref.~\onlinecite{nandini}, for example, for a discussion about the presence of an 
intermediate metallic phase when disorder increases). 
The Mott insulator is characterized by a gap in the typical DOS ($\rho_{typ}(0)=0$) and a 
finite frequency integrated typical-DOS $N$ (see panel a), as well as a single occupation of all 
sites (see panel b). The metallic phase, on the other hand, features a quasi-particle peak in 
the typical DOS, nonzero integrated DOS $N$, and a variable site occupation $n_i$. 
Finally, the correlated Anderson insulator shows a vanishing typical DOS, indicating that 
all the states are localized and as such do not contribute with spectral weight to 
the typical DOS.\cite{TMT,Vollhardt}
For this reason, the frequency integrated typical DOS goes to zero when the system
approaches the Anderson insulator, and thus can be used as an order parameter 
that signalizes this transition.
Furthermore, within the TMT-DMFT the Anderson insulating phase corresponds to a two-fluid 
phase:\cite{Carol2} it consists of empty and doubly occupied sites, characteristic 
of non-interacting Anderson insulators, as well as singly occupied sites, characterizing Mott localized states (see the results for $W=4$ in panel b).

We find good agreement between our diagram and others known in the literature at 
$T=0$.\cite{Vollhardt, Carol2} The most relevant effects of finite, but small temperature are over the Mott coexistence region, which spans over a smaller range of $U$ in comparison  with the $T=0$ case. The real axis IPT impurity solver makes it possible 
to solve TMT-DMFT equations for a broad range of parameters and several temperatures. 
In the following, we concentrate on the range of parameters near the phase transition, 
and, in particular, near the coexistence region of metallic and insulating solutions.

\section{Mott transition for weak and moderate disorder $W< 2B$}\label{IV}

In this Section we analyze the coexistence region for weak and moderate disorder, which corresponds to $W<W^*$, $W^*\approx 1.7$. At this regime, the critical $U$ for the Mott transition is greater than the disorder strength. Although the phase transition described within TMT-DMFT is qualitatively similar as that of CPA-DMFT, some Anderson localization effects are already observed.

\subsection{Coexistence region}
\begin{figure}[b]
\begin{center}
\includegraphics[width=\linewidth]{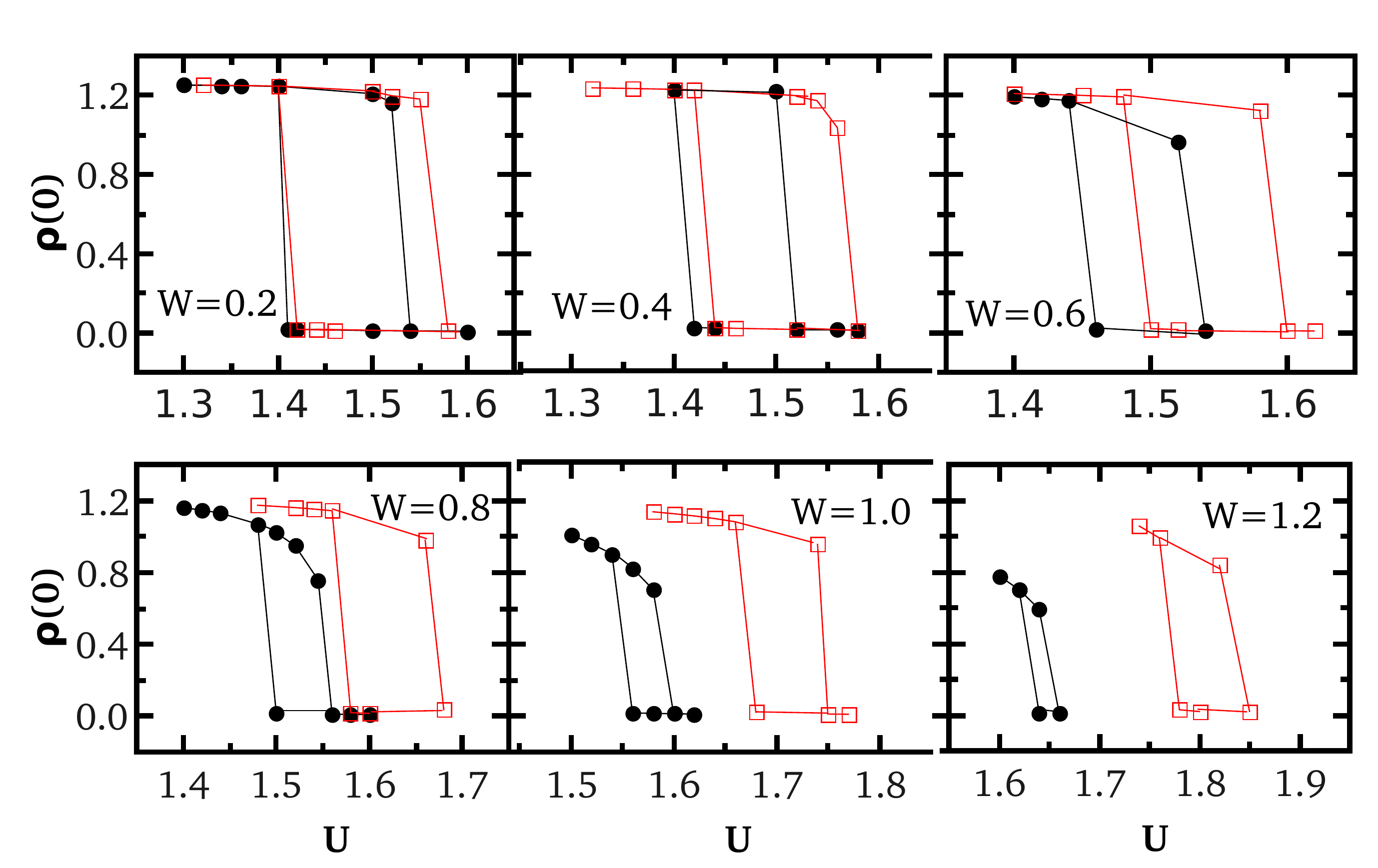}
\caption{(Color online)
Hysteresis curves for the DOS at the Fermi level obtained by increasing and 
decreasing $U$ at a fixed temperature $T=0.01$. The curves enclose the coexistence region. 
The open squares were obtained within CPA-DMFT, while the filled circles correspond to TMT-DMFT
results.}
  \label{fig6}
  \end{center}
\end{figure}

To obtain the coexistence region within CPA-DMFT or TMT-DMFT, for a fixed temperature 
$T<T_c$, we start from a metallic initial bath and increase $U$ to find $U_{c2}$, 
which corresponds to the interaction value at which $\rho(0)$ goes to zero, 
indicating the disappearance of the quasi-particle peak in the DOS. Alternatively, when starting 
from an insulator, by decreasing $U$ we find $U_{c1}$ where $\rho(0)$ becomes 
finite, indicating the closure of the gap at the Fermi level. This procedure 
allows us to obtain hysteresis curves of $\rho(0)$ as a function of $U$, 
which enclose the coexistence region (see Fig.~\ref{fig6} for examples of these
hysteresis curves). 
For a given $W$, we can repeat this procedure for different 
temperatures and determine the two spinodal lines, $U_{c1}(T)$ and $U_{c2}(T)$, 
defining the coexistence region. The temperature at which the two spinodal lines 
merge gives the critical temperature, $T_c$, which corresponds to a second
order critical end point. 

\begin{figure}
\includegraphics[scale=0.35]{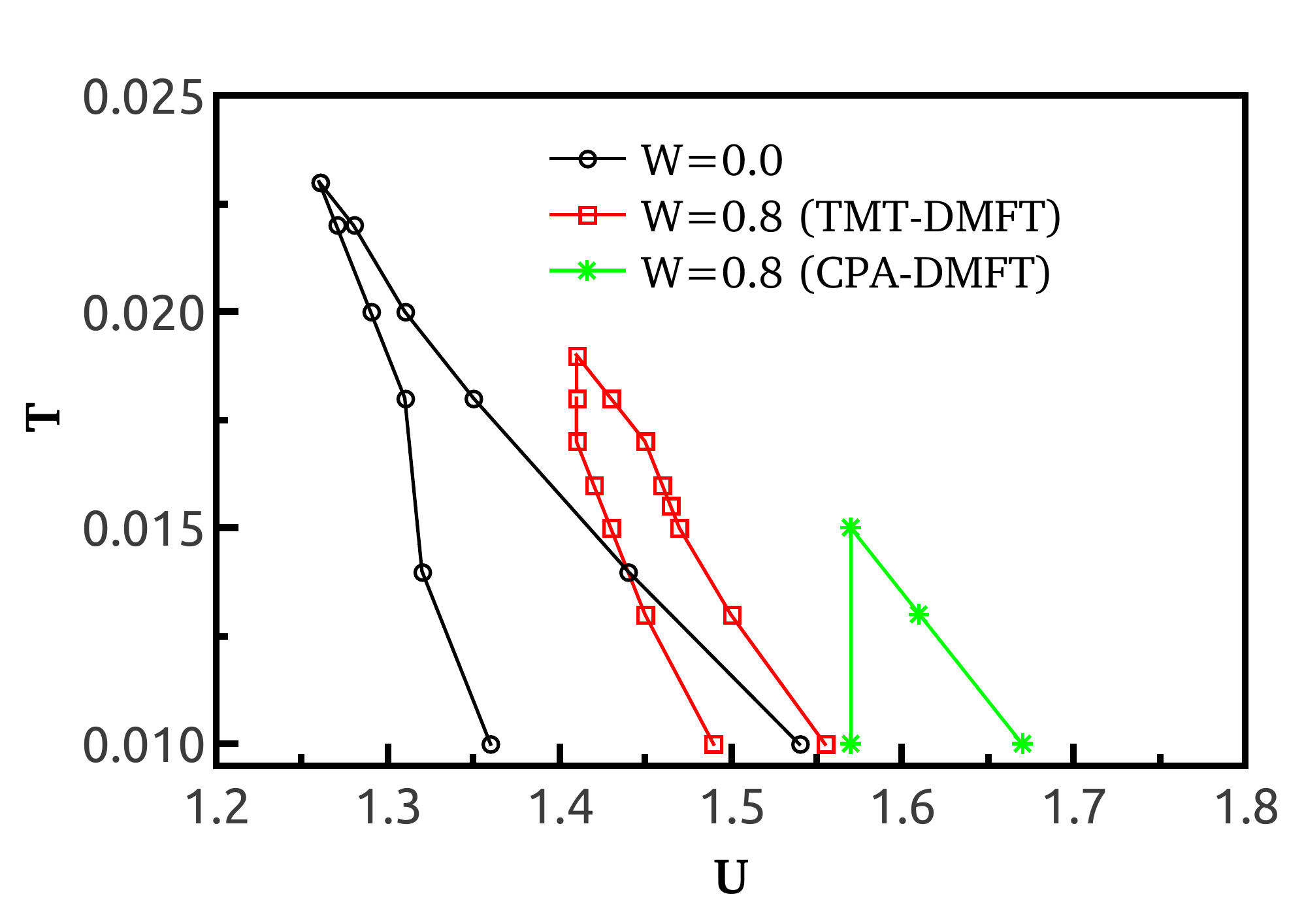}
\caption{(Color online) Spinodal lines enclosing the coexistence region for 
the clean system ($W=0$) and the disordered case ($W=0.8$) obtained within 
both TMT-DMFT and CPA-DMFT.} 
\label{fig3} 
\end{figure}

\begin{figure}
\includegraphics[width=\linewidth]{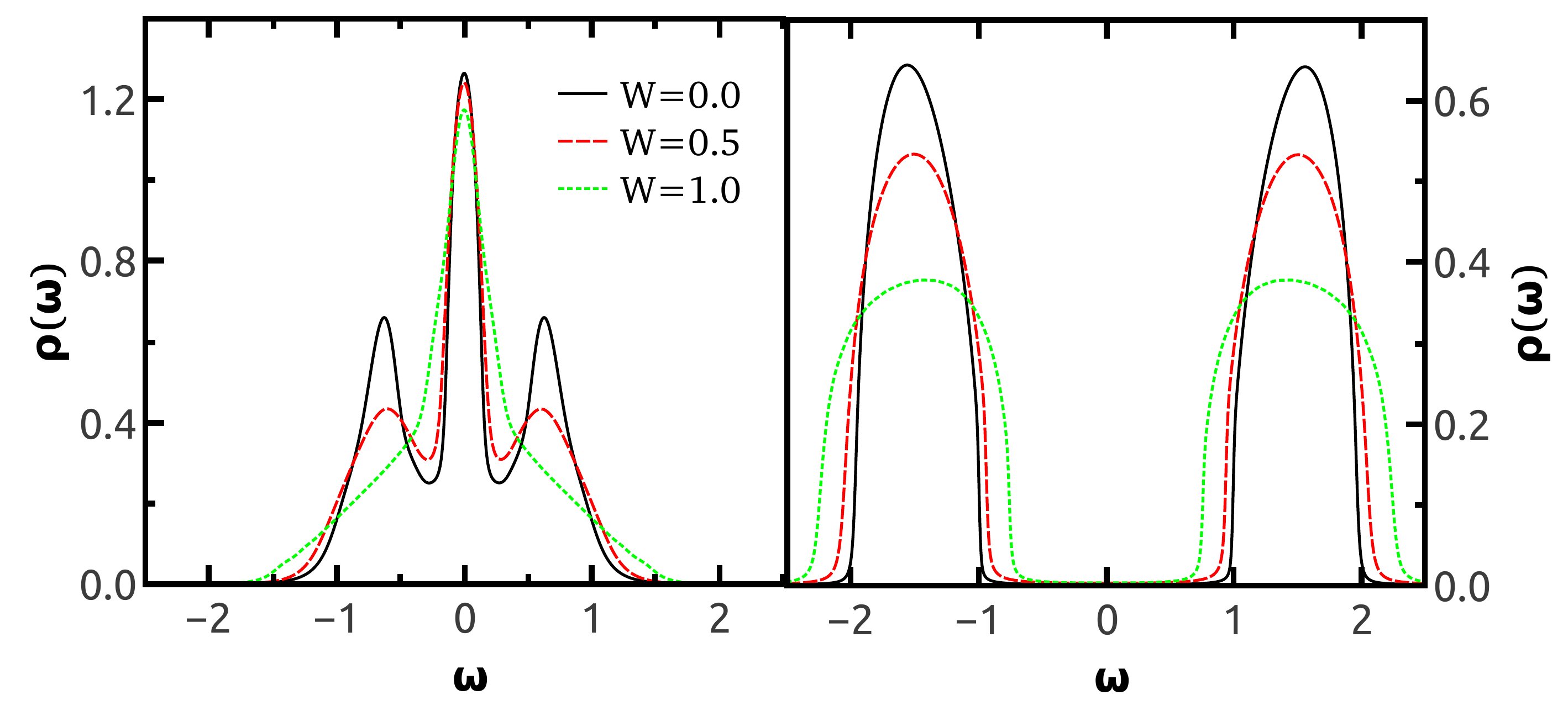}
\caption{(Color online) Average DOS obtained within CPA-DMFT for different values of 
disorder at fixed temperature $T=0.01$. Disorder broadens the bands in both the metallic 
(left panel, for $U=1$) and the insulating (right panel, for $U=3$) phase.} 
\label{fig4} 
\end{figure}

Figure  ~\ref{fig3} shows the coexistence region obtained as described above for the clean 
case ($W=0$) and for a disordered system ($W=0.8$), both within TMT-DMFT and CPA-DMFT.  
According to our results, when disorder is added to the system the 
critical $U$ at which the transition occurs increases in comparison with the clean
case. This happens because the general effect of disorder is to broaden the bands, as shown
in Fig.~\ref{fig4}, when the CPA-DMFT calculation is performed both inside the metallic 
and the insulating phase. Another general effect of disorder seen in
the results of Fig.~\ref{fig3} is that the temperature of the second order critical point 
decreases with disorder, in agreement with previous CPA-DMFT calculations.\cite{Carol1} These general consequences of disorder do not depend on the inclusion of Anderson localization effects, since they are observed even within CPA-DMFT approach.

\begin{figure}
\begin{center}
\includegraphics[width=\linewidth]{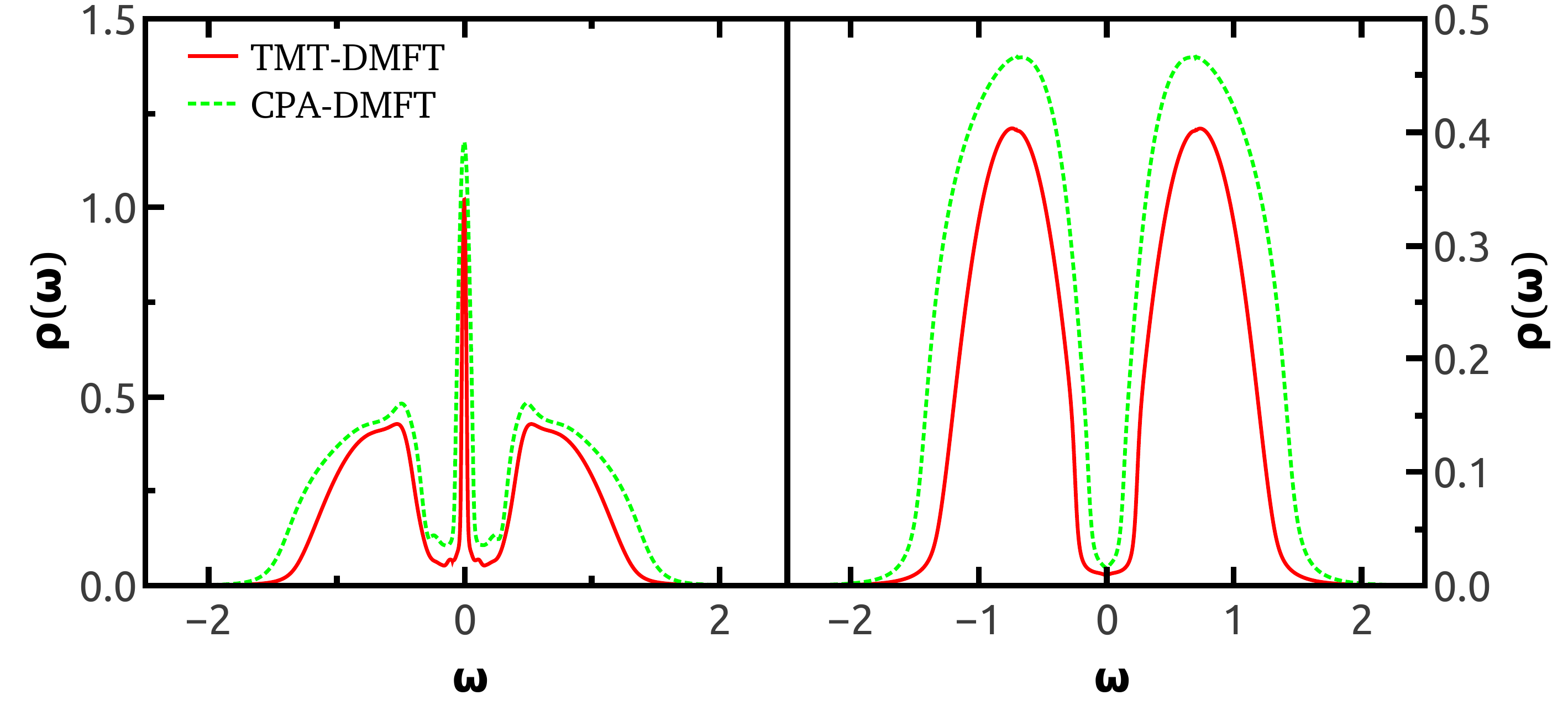}
\caption{(Color online) 
Typical (TMT-DMFT) and average (CPA-DMFT) values of the DOS as a function of frequency
showing that Anderson localization effects start at the band edges, since both localized and 
extended states contribute to the average DOS, while only extended ones contribute to the 
typical DOS. Left panel shows results for $U=1.5$, while those in the right panel are for 
$U=1.6$, both at $T=0.01$.}
\label{fig5}
\end{center}
\end{figure}

To carefully study the effects of Anderson localization we compare the results obtained 
within TMT-DMFT with those of CPA-DMFT. As can be seen in Fig.~\ref{fig3} for $W=0.8$, the critical $U$ at which the transition occurs is smaller within TMT-DMFT than within CPA-DMFT. Moreover, a narrower coexistence region is observed within the former. To understand these results,  one should consider that the wave function localization starts at the band edges and that localized states do not contribute with spectral weight to the typical DOS. For these reasons, in the presence of Anderson localization narrower bands are observed in comparison with CPA-DMFT results, both in the metallic and the insulating phase, as can be seen in Fig.~\ref{fig5}. This is the opposite effect as that described in the previous paragraph regarding the effects of adding disorder to a clean system. As a consequence, the coexistence region within TMT-DMFT is seen in between that of a clean system and that obtained within CPA-DMFT for the same value of disorder.

From the TMT-DMFT and the CPA-DMFT hysteresis curves shown in Fig.~\ref{fig6} we see that the Anderson localization effects over the coexistence region become more important as the disorder increases.
As $W$ approaches $W^* \approx 1.7$, the width of the TMT-DMFT coexistence region vanishes and we were not able to observe the hysteresis even at the lowest temperatures $T=0.005$ (see section \ref{V}).
In contrast, in the CPA-DMFT solution,\cite{Carol1} the coexistence region with finite small $T_c$ is observed even for very large $W$.

\subsection{Crossover regime and the critical temperature $T_c$} 

As seen in Fig.~\ref{fig6}, the coexistence region shrinks as disorder increases,
making it difficult to obtain the critical temperature $T_c$ from the merging of
the two spinodal lines. One alternative is to determine $T_c$ from the results obtained
above it, that is, in the crossover region between metal and insulator. This was shown to be possible in the clean case and in the present work we extend this analysis to the disordered system. 

The quantum Widom line (QWL) associated with the Mott transition is defined in Refs.~\onlinecite{Hanna,Jaksa,Jaksa2015,Kanoda} in analogy with the classical Widom line\cite{Simeoni2010WL} as the instability (crossover) line above the critical end-point $(U_c,T_c)$. 
It starts at the critical end-point and goes to higher temperatures (above the coexistence region) as a continuation of the first-order phase transition line. It is associated with the (zero temperature) quantum critical point, which is masked by the coexistence region in the case of the Mott transition. The QWL can be defined from the free energy functional $F_L[G(i\omega_n)]$ and can be used to determine $T_c$ from the behavior at higher temperatures, as explained (for the clean case) in Refs.~\onlinecite{Hanna, Jaksa}.
With the objective of applying the QWL analysis to obtain $T_c$ in the disordered case, 
here we review this procedure. 

The Landau free energy functional of the Hubbard model as a functional of $G(i\omega_n)$ is given by 
 
$$
F_L[G(i \omega_n)] = -T t^2 \sum_n G^2(i \omega _n) + 
F_{imp}[G(i\omega _n)],
$$
where the first term represents the energy needed to form the bath around a given 
site and the second term describes the energy of the electron at the impurity level 
surrounded by the bath, that is, the free energy of the single-impurity problem. 
The DMFT (TMT-DMFT) equations are obtained by minimizing $F_L[G(i \omega_n)]$ with respect to $G(i\omega_n)$.

The curvature $\lambda$ of the above free energy functional with respect to $U$ is finite and minimal along the crossover line and is zero at the second order critical point. This curvature can be identified with the convergence rate of the iterative DMFT calculation,\cite{Hanna, Jaksa} that is, $\lambda(U,T)$ corresponds to the slope of the convergence rate $\ln \{\mbox{Im} [G^{(it)}(0)-G^{(it-1)}(0)] \}$ as a function of the step $it$ of the iterative calculation. 
Repeating the calculation for different values of $T$, we obtain the curve 
$\lambda_{min}=\lambda(T)|_{U^*}$, where $U^*$ is the point at which $\lambda$ is minimum for a given $T$. This line can be extrapolated, to $\lambda_{min}|_{T=T_c}=0$, since the curvature of the free energy functional is zero at the second order critical point.   
 
\begin{figure}[h]
\begin{center}
\includegraphics[width=\linewidth]{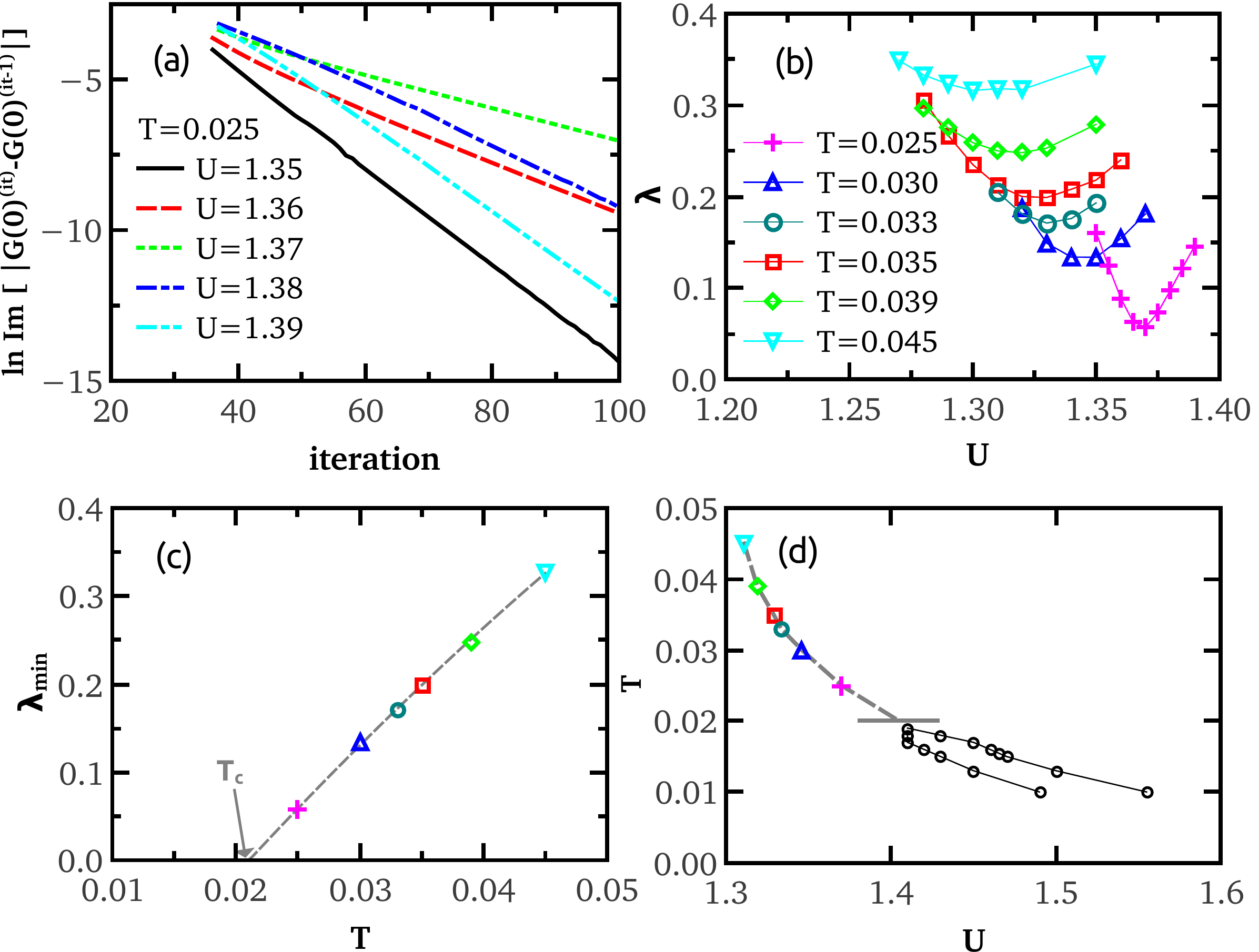}
\caption{(Color online) QWL analysis for the disordered system with $W=0.8$ 
described by TMT-DMFT. See the text for the explanation of the results in each 
panel.}
\label{fig7} 
\end{center}
\end{figure}

The procedure is illustrated in Fig.~\ref{fig7} 
for the disordered system with $W=0.8$. For each value of $U$, we obtain the free energy 
curvature $\lambda$ from the convergence rate of the typical Green's function through the 
iterative steps, as presented in (a) for $T=0.025$. For a fixed temperature and 
different values of $U$, we obtain the corresponding $\lambda(U)|_T$ curve. 
Repeating this procedure for different temperatures, we obtain the set of curves 
$\lambda(U)|_T$ presented in Fig.~\ref{fig7}b.
The minima $\lambda_{min}$ of these curves are shown in panel (c), and
we obtain $T_c$ as the temperature at which $\lambda_{min}=0$. 
Finally, panel (d) shows the crossover line obtained from data in panel (b),
$T_c$ obtained through the QWL analysis (gray horizontal line), and the two spinodal lines. 
We conclude that the $T_c$ calculated from the QWL analysis coincides with the $T_c$ obtained from 
the merging of the two spinodal lines that define the coexistence region.

\begin{figure}
\begin{center}
\includegraphics[width=2.8 in]{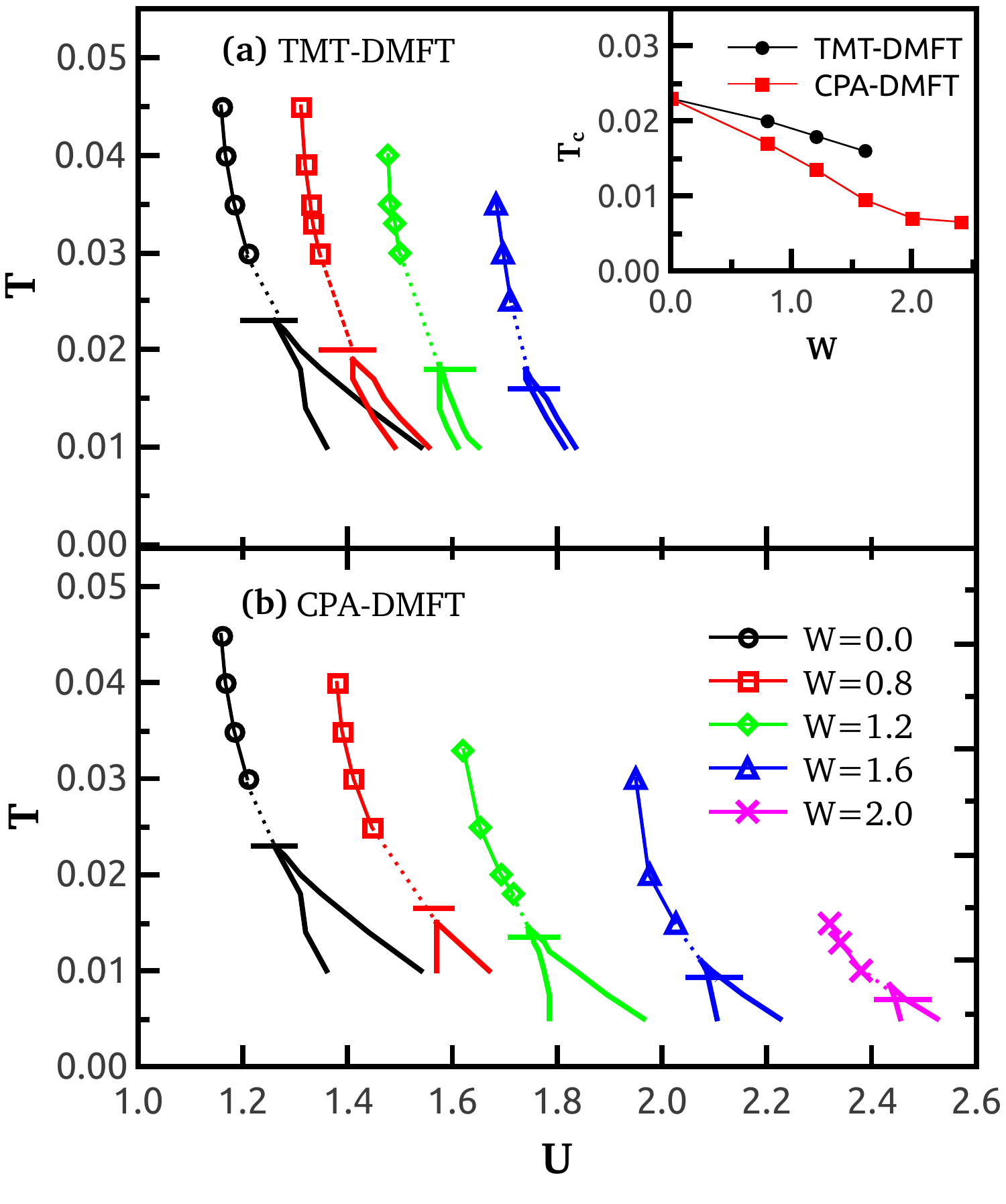}
\caption{(Color online) QWL and coexistence regions obtained within TMT-DMFT (a) and CPA-DMFT (b) for 
different values of disorder (CPA-DMFT coexistence regions for $W\geq1.2$ were obtained from Ref.~\onlinecite{Carol1}). 
The horizontal lines represent $T_c$ obtained from 
the corresponding QWL, calculated as exemplified in Fig.~\ref{fig7} (c). The inset shows 
these $T_c$ values as a function of disorder.}
\label{fig8} 
\end{center}
\end{figure}

In Fig.~\ref{fig8} we show the QWL and the critical temperatures obtained from them 
as we vary the system disorder, both within TMT-DMFT 
and CPA-DMFT. 
For disorder strengths $W\gtrsim 1.6$, we find a nonlinear
behavior of the TMT-DMFT convergence rate as a function of the iteration step; we were
thus unable to use the QWL analysis discussed to evaluate $T_c$ for very large disorder. For $W<1.7$,
both methods predict that $T_c$ decreases when $W$ 
increases (see also the inset in Fig.~\ref{fig8}a). The critical temperature $T_c$ is higher within 
TMT-DMFT than within CPA-DMFT, although the 
coexistence region becomes (very) narrow in the presence of Anderson localization 
effects (TMT-DMFT results). However, $T_c$ always remains finite within CPA-DMFT even for very 
large disorder strength\cite{Carol1}, whereas we do not observe the coexistence region for $W\gtrsim 1.7$ in TMT-DMFT (see next section).
Our numerical TMT-DMFT 
solution indicates that the $T_c$ abruptly drops to zero as the coexistence region disappears for $W\approx 1.7$.

\section{Mott-Anderson transition for strong disorder $W \gtrsim 2B$ }\label{V}

\begin{figure}
\begin{center}
\includegraphics[width=2.8 in]{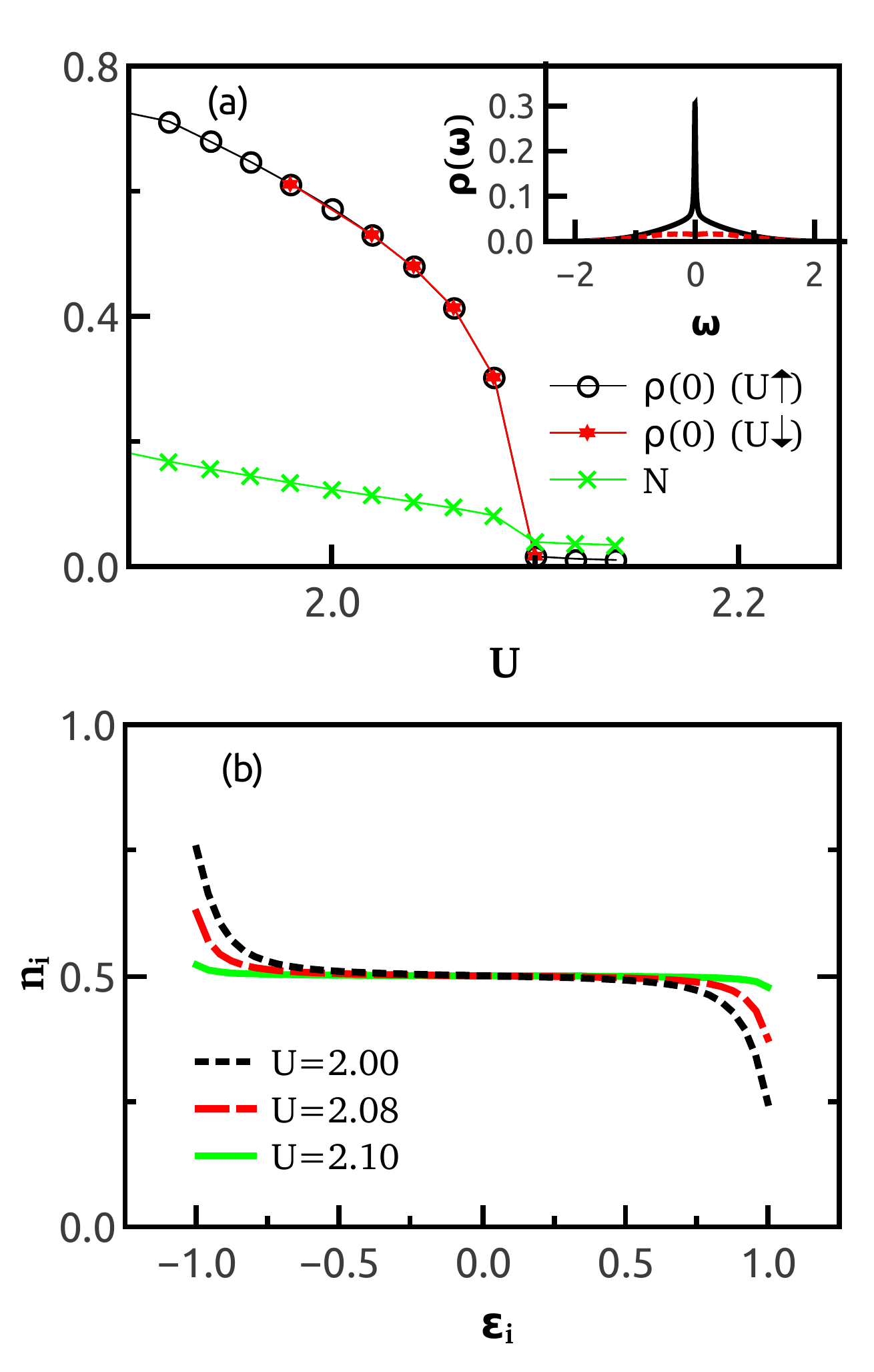}
\caption{(Color online)
Results obtained within TMT-DMFT for $W=2.0$ at $T=0.01$. 
Panel (a) presents the typical DOS at the Fermi level obtained by increasing $U$ 
(black circles) and decreasing $U$ (red stars); no coexistence region 
is observed. In the same panel we can also see the frequency-integrated typical DOS 
$N$ as a function of $U$. The inset shows the typical DOS as a function 
of frequency for $U=2.08$ (black solid line) and $U=2.10$ (red dashed line). Finally, panel 
(b) presents the occupation number per spin as a function of the site energy as the 
transition is approached.}
\label{fig9} 
\end{center}
\end{figure}

Within the TMT-DMFT calculation, as we increase disorder, the value of the critical 
$U$ becomes closer to the disorder
width $W$. 
For $U \sim W \sim 2B$ both Mott and Anderson routes to localization become equally relevant,
and it becomes the most difficult to precisely understand the mechanism of the MIT.
In Fig.~\ref{fig9} we show the results for  $W=2.0$  at $T=0.01$. The transition is seen to take place at 
$U \approx 2.09$. Moreover, if we look at the results for the typical DOS at the Fermi 
level when $U$ increases, as well as when $U$ decreases (see panel (a)), we observe no 
hysteresis, even if we decrease the temperature down to $T=0.005$, in contrast to the 
results shown in Fig.~\ref{fig6}. Since $\rho_{typ}(0)$ becomes zero, the system certainly 
goes through a MIT - but to what type of insulator does the system go to? 

To answer the question above, we first look at the frequency integrated typical DOS 
$N$, which can be considered an order parameter in the case of the Anderson 
transition, as discussed in the beginning of the paper. As 
can be seen in Fig.~\ref{fig9}(a), $N$ becomes very small but is still
finite when $\rho_{typ}(0) \rightarrow 0$, suggesting that the transition is {\it not} of 
the Anderson type. The nature of the transition can finally be confirmed by 
analyzing the occupation number per spin $n_i$ as a function of the site energy 
close to the transition, which can be seen in panel (b). As $U$ increases towards 
the MIT, all sites become singly occupied, which is a characteristic of the Mott insulator. 
Although of the Mott type, the Hubbard subbands are strongly reduced for this 
value of $W$, as can be seen in the DOS presented in the inset,
which is consistent with our expectation that both Mott and Anderson routes to 
localization are relevant in this regime of $U \approx W$.

Interestingly, our analysis of Fig.~\ref{fig9} suggests 
that for $W = 2.0$ there exists a transition between a metal and a Mott insulator 
in the {\it absence} of a coexistence region. Indeed, according to the phase diagram
(Fig.~\ref{fig1}), the same behavior is observed in a small range around
$U \approx W \approx 2$. According to Figs.~\ref{fig6} and ~\ref{fig8}, TMT-DMFT predicts 
the coexistence region to become (very) narrow when the system is in the $U < W$ regime 
and disorder increases. When the system enters the $U \sim W$ regime, the two spinodal 
lines seem to merge and no coexistence is observed, suggesting that $T_c$ abruptly goes to zero 
due to the Anderson localization effects. Our results are in general agreement with 
the $T=0$ phase diagram of Ref.~\onlinecite{Vollhardt}, while presenting much more detailed
analysis of the MIT with the vanishing coexistence region.

For $W \gtrsim 2.3$, one can find a direct crossover between the two insulators, Mott 
and correlated Anderson, without an intermediate metallic phase; this crossover is represented 
by red squares in our diagram of Fig.~\ref{fig1}. 
To distinguish between the two insulators, we have looked at the occupation number as a 
function of site energy, as exemplified in Fig.~\ref{fig2}. Our results show that when $W<U$ 
all the sites are single occupied, characterizing a Mott insulator; when $W>U$, on the
other hand, there are sites with energy larger than $U/2$, which are empty, sites with energy 
smaller than $-U/2$, which have double occupancy, and also sites occupied with one electron, 
characterizing the two fluid behavior of the correlated Anderson insulator. According to
these results, as might have been expected from the two fluid picture of the Mott-Anderson 
insulator,\cite{Carol2} the crossover between the two insulators is seen to take place at 
$W \approx U$.

\section{Conclusions}\label{VI} 

In this work, we studied Mott and Anderson routes to localization by using a 
combination of dynamical mean field theory (DMFT) and typical medium theory (TMT) to 
solve the disordered Hubbard model. According to our TMT-DMFT results, Anderson 
localization has important effects near the Mott transition, specially on the 
coexistence region of metallic and insulating phase that exists below a critical temperature
$T_c$. In the presence of small and moderate disorder $W$, the TMT-DMFT transition is
qualitatively similar as in the CPA-DMFT case (which does not describe localization due to disorder), 
and the main precursors of the Anderson localization
are seen in the narrowing of the coexistence region in comparison with CPA-DMFT.
As the disorder further increases, for $W \gtrsim 2B$ (where $B$ is the bandwidth for $U=W=0$), 
the transition occurs at $U \approx W$ and our results indicate that Anderson and 
Mott routes to localization become equally important. The critical temperature $T_c$
abruptly goes to zero for $W = W^* \approx 1.7 B$. For $1.7 B < W \sim U < 2.3 B$ the typical DOS
at the metal-insulator transition is strongly reduced, but the states are nearly half-filled
irrespective of the on-site energy, indicating dominantly Mott character of the MIT, although no coexistence region is observed.
For even larger disorder, $W > 2.3$, there is a crossover
between the Mott and the correlated Anderson insulator.

The observation of a Mott transition without a coexistence region suggests that
the nature of the transition has changed from first to second order as disorder
increases. For the clean system, it has been shown\cite{Hanna} that at $T$ just
above $T_c$ the resistivity as a function of temperature shows a scaling behavior, 
which is compatible with an assumption of quantum criticality.
In other words, despite the presence of a coexistence region between the metallic 
and the Mott insulating phase at small temperatures, at intermediate 
temperatures the system seems to be controlled by a hidden quantum critical point.
Very recently an experimental work on $\kappa$-organics has confirmed the presence of this quantum 
critical regime at intermediate temperatures.\cite{Kanoda} 
In this respect, it will be very important to compare the TMT-DMFT phase diagram 
and charge transport with the
experiments on disordered correlated systems. Preliminary results,\cite{Kanoda_privcomm}
on introducing disorder by X-ray irradiation, show that $U_c$ indeed increases with disorder, 
while $T_c$ also decreases, and seems to vanish at some finite disorder.

We acknowledge the HPC facility at Florida State University and CENAPAD-SP, where 
part of the results were obtained. This work was supported by the Brazilian agencies
CNPq, FAPEMIG and Capes (H.~B.~and M.~C.~O.~A.), 
by the Ministry of Education, Science, and Technological Development of the Republic of 
Serbia under project ON171017 (J.~V. and D.~T.),
and by the NSF, Grants No. DMR-1005751 and No. DMR-1410132 (V.~D.).

\end{document}